\documentclass{article}  
\usepackage{breckenridge}
\usepackage{epsf}
\frompage{000} \topage{000}                                              
\def\rightmark{$\phi$ Production at $\sqrt{s_{NN}}$ = 200 GeV}
\def\leftmark{Jingguo Ma}
 
\title{Mid-rapidity $\phi$ meson production at $\sqrt{s_{NN}}$ = 200 GeV 
Au+Au and pp collisions from STAR} 
\authors{
{Jingguo Ma$^{1,a}$ for the STAR Collaboration %
}\\[2.812mm]
{\normalsize
\hspace*{-8pt}$^1$ University of California, Los Angeles, \\ 
90095 Los Angeles, USA\\[0.2ex] 
}}

\abstract{We present the results for the measurement of $\phi$ meson 
production in $\sqrt{s_{NN}}$ = 200 GeV Au+Au and pp collisions at the 
Relativistic Heavy Ion Collider (RHIC). Using the event mixing 
technique, spectra and yields are obtained from the
$\phi\rightarrow K^{+}K^{-}$ decay channel for five centrality bins in
Au+Au collisions and in pp collisions. We observe that the spectrum 
shape in Au+Au collisions depends weakly on the centrality and the shape
of the spectrum in pp collisions is significantly different from that
in Au+Au collisions.}

\PACS{{\tt 25.75.Dw}}
 
\begin{document}
 
\maketitle
\setcounter{page}{1}

\section{Introduction}\label{intro}

Strangeness production in Relativistic Heavy Ion Collisions may
provide detailed information on the collision dynamics \cite{bib1}. 
The enhanced production of strangeness in nucleus-nucleus collisions
is predicted to be a signature that the collisions go through a
deconfined stage - namely the quark gluon plasma (QGP)
\cite{bib2,bib3}. The $s\bar{s}$ quark content for the $\phi$ meson is
of particular interest: The production of $\phi$ meson may be sensitive
to strangeness enhancement \cite{bib2,bib3,bib4}. The possible mass
and decay width modification of $\phi$ meson due to the expected 
partial chiral symetry restoration and in medium effect in the hot and
dense matter has been a topic of both theoretical and experimental
investigations \cite{bib5,bib6,bib7,bib8,bib9}. The $\phi$ meson may
retain the information from the early partonic stage
of the collisions since it is expected that $\phi$ meson interacts
weakly with nonstrange hadrons during the hadronic stage
\cite{bib3,bib4}. Finally, the comparison of the production of $\phi$
meson in Au+Au and pp collisions at the same beam energy may yield
knowledge on the production mechanisms of $\phi$ meson and the
evolution dynamics of the colliding system.

\begin{figure}[htb] 
\vspace*{-.1cm}
\begin{center}
\leavevmode \epsfysize=5.0cm 
\epsfbox{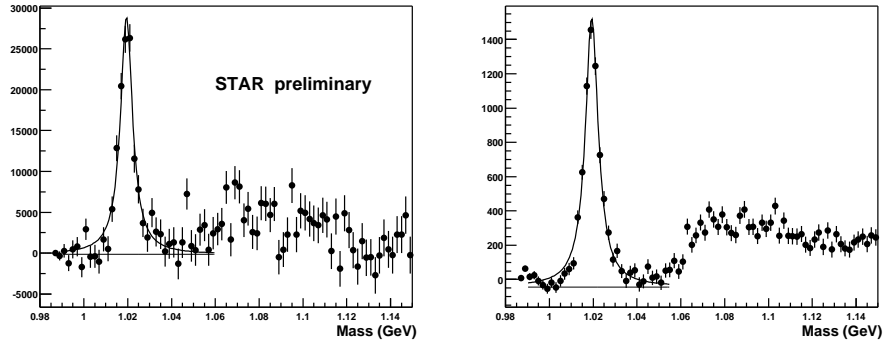}
\end{center}
\vspace*{-.8cm}
\caption[]{Background subtracted $K^+$$K^-$ pair invariant mass
distribution from the top 10\% central Au+Au collisions (left pannel)
and pp collisions (right pannel). The $\phi$ peak is fitted to a 
Breit-Wigner function plus a linear background function representing
the background. The residual background at higher mass region comes
from misidentified $K_s^0$ decay products.}
\label{fig1}
\end{figure}

\section{Experiment and Data Analysis}\label{techno} 

The results on $\phi$ meson production at $\sqrt{s_{NN}}$ = 130
GeV from the STAR detector have been reported \cite{bib10}. The data
presented here was taken by the STAR detector \cite{bib11} 
during the second run of RHIC collider in the year 2001. The main
components of the STAR detector used in this analysis are a large
acceptance Time Projection Chamber (TPC) \cite{bib12}, a Central
Trigger Barrel (CTB) and two Zero Degree Calorimeters. The TPC is 
placed in an uniform magnetic field as the tracking device for
charged particles. The CTB and ZDC are used for triggering. In this
analysis about 2.1M events from minimum bias trigger data and 0.5M
events from central trigger data in Au+Au collisions and 4M events
from minimum bias trigger data in pp collisions are used after all
event selection cuts \cite{bib10}.

  The particle identification is achieved by correlating the measured
energy loss ($dE/dx$) of the particle in the TPC gas with its
momentum. In this analysis, a track is selected as a kaon candidate as
long as its $dE/dx$ is within $2\sigma$ of the kaon Bethe-Bloch
curve. Due to limited resolusion of the detector, the pion and kaon 
dE/dx bands merge at momentum greater than 0.7 GeV, resulting in the
pion contamination in the kaon candidates. The centrality in Au+Au
collisions is defined by the fraction of the total inelastic hadronic 
cross-section, i.e. by dividing the raw charged hadron multiplicity
distribution into centrality classes. For the $\phi$ analysis, the 
centrality division for the Au+Au collisions is: top 5\% from central
trigger data and top 10\%, 10-30\%, 30-50\%, 50-80\% from minimum bias
trigger data.

  The $\phi$ signal is built by calculating the invariant mass of
every selected $K^+$$K^-$ pair. The shape of the combinatorial
background is calculated by event mixing technique
\cite{bib13,bib14}. For the Au+Au data analysis, each event is
mixed with another event in the same centrality class, while each
event is mixed with four other events in the pp data analysis in
order to get a better description of the background with higher
statistics.

\section{Results}\label{others}
The background subtracted $K^+$$K^-$ pair invariant mass distribution
averaged over 0.4 GeV/c $\leq p_T \leq$ 3.9 GeV/c is shown in fig.
\ref{fig1}. The $\phi$ peak is fit to a Breit-Wigner function plus a
linear function representing the background in each $p_T$ bin. The
measured mass for the  $\phi$ meson in both Au+Au and pp collisions 
is $1019\pm0.7$ MeV/$c^2$, which is consistent with the $\phi$ meson mass 
from the Particle Data Group \cite{bib15}. The measured width is also consistent 
with the $\phi$ natural width convoluted with the resolution of the 
STAR detector. The residual background at mass region greater than
1.06 GeV/$c^2$ mainly comes from $K_s^0$ decay products when pions are 
misidentified as kaons. A small fraction of it may also come from 
$\Lambda$ decays where both of its decay daughters are misidentified
as kaons.
 

\begin{figure}[htb] 
\vspace*{-.1cm}
\begin{center}
\leavevmode \epsfysize=6.0cm 
\epsfbox{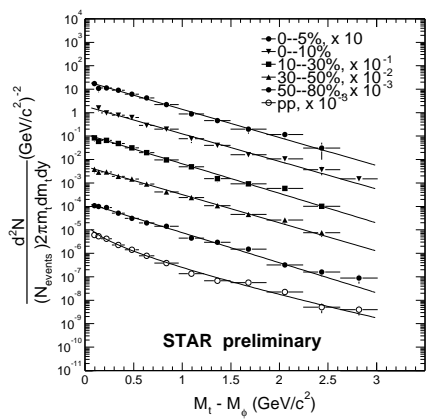}
\end{center}
\vspace*{-.8cm}
\caption[]{$\phi$ invariant multiplicity distribution as a function of
transverse mass for five centrality bins in Au+Au collisions (filled
symbols) and pp collisions (open symbols) at $\sqrt{s_{NN}}$ = 200
GeV. The spectra are scaled by different factors to guide eyes.}
\label{fig2}
\end{figure}

\begin{figure}[htb] 
\begin{center}
\leavevmode \epsfysize=6.0cm 
\epsfbox{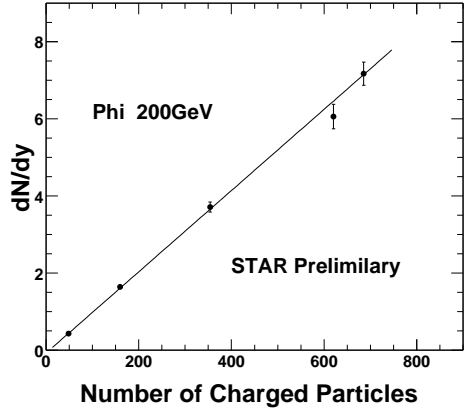}
\end{center}
\caption[]{$\phi$ dN/dy as a function of number of charged
particles for Au+Au collisions. dN/dy of $\phi$ scales linearly
with number of charged particles in Au+Au collisions.}
\label{fig3}
\end{figure}

\begin{figure}[htb] 
\begin{center}
\leavevmode \epsfysize=6.0cm 
\epsfbox{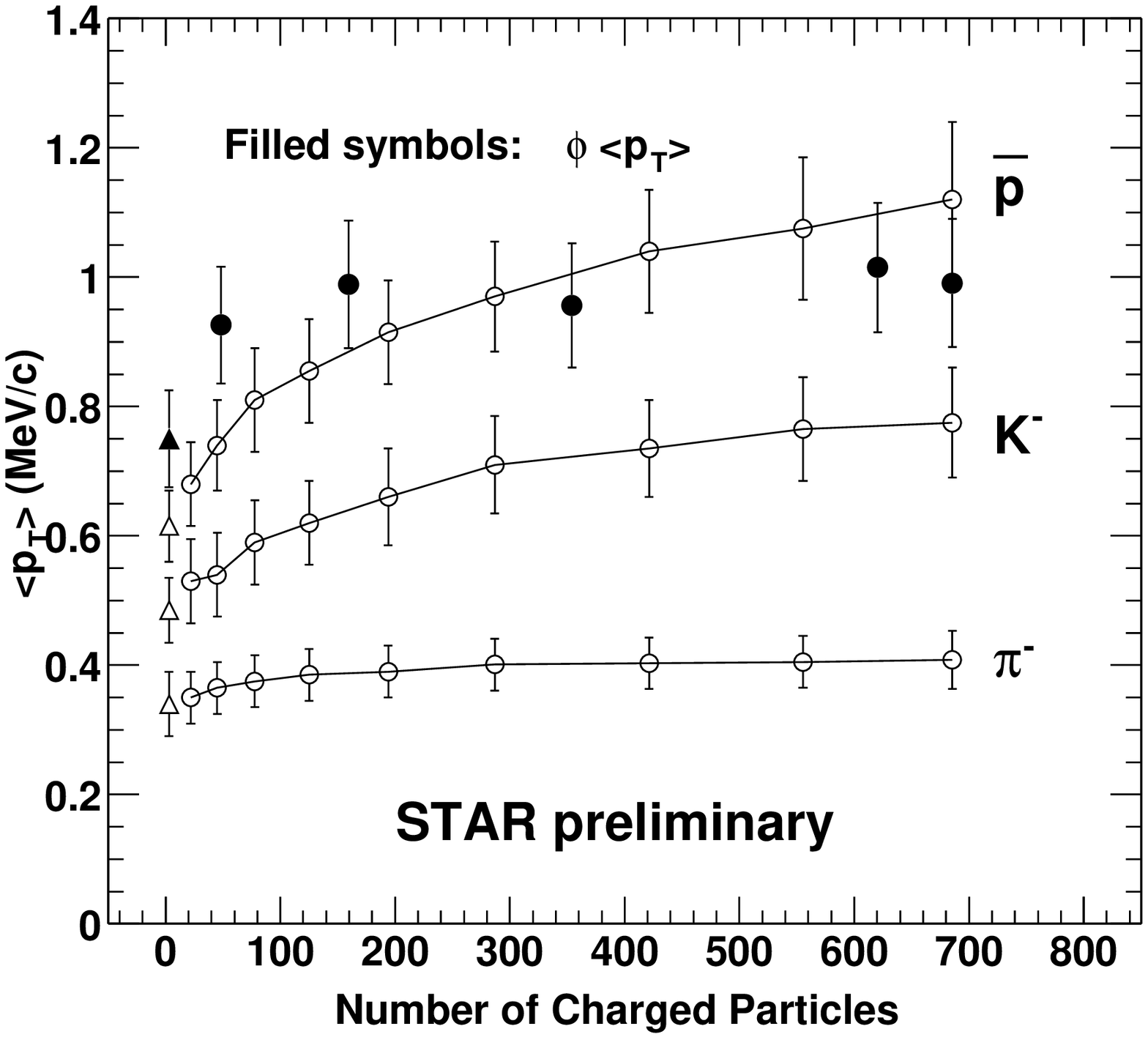}
\end{center}
\caption[]{The $<p_T>$ of $\phi$, $\pi^-$, $K^-$ and $\bar{p}$ at
$\sqrt{s_{NN}}$ = 200 GeV Au+Au and pp collisions as a function of
the number of charged particles. The triangles represent the $<p_T>$ 
obtained in pp interactions and the circles correspond to the $<p_T>$ 
measured in Au+Au collisions for different centralities. $\phi$
$<p_T>$ in pp collisions is extracted from power law $p_T$ fit
function. In Au+Au collisions, $\phi$ $<p_T>$ is extracted from
exponential $m_T$ fit function. $\phi$ $<p_T>$ shows a very different
behaviour when compared to $\pi^-$, $K^-$ and $\bar{p}$.}
\label{fig4}
\end{figure}

  In order to obtain the resonance yield, detector acceptance and
efficiency corrections were applied to the uncorrected number of
$\phi$ for each centrality and $p_T$ bin. The acceptance and efficiency
corrections were done by embedding simulated kaons from $\phi$ decays
into real events using GEANT, and by passing them through the full
reconstruction chain \cite{bib10,bib16}. The corrected $\phi$ invariant
multiplicity distributions at mid-rapidity ($|y| <$ 0.5) as a function
of $m_T - m_{\phi}$ are depicted in Fig. \ref{fig2}, where
$m_{\phi}$ = 1019.4 MeV/c$^2$ is the average $\phi$ mass reported in
\cite{bib15}. The transverse momentum coverage for this measurement is
0.4 $\leq p_T \leq$ 3.9 GeV/$c$, which corresponds to 85$\%$ of the
$\phi$ yield at mid-rapidity. In Au+Au collisions, the spectra are
fitted by an exponential function in $m_T - m_{\phi}$ for all
centrality bins. In pp collisions, the spectrum is better represented
by a power law function in $p_T$. Thus the spectra shape changes from
pp to Au+Au collisions. In Au+Au collisions, the inverse slopes
extracted from the exponential function fit to the spectra do not
depend strongly on the collision centrality.

  The $\phi$ yield ($dN/dy$) obtained from the fit to the spectra in
Au+Au collisions as a function of the number of charged particles is 
shown in Fig. \ref{fig3}. The number of charged particles
corresponds to the acceptance and tracking efficiency corrected
charged particle multiplicity within $|\eta| <$ 0.5, where $\eta$ is 
the pseudorapidity. The measured $\phi$ yield scales linearly with the
number of charged particles produced in Au+Au collisions.

  The $\phi$ $<p_T>$ calculated from the fits to the spectra as a
function of the number of charged particles is depicted in Fig.
\ref{fig4}, where the solid triangle and the solid circles
sketch the measurements in pp and Au+Au collisions, respectively. In
this figure, the $<p_T>$ of $\pi^-$, K$^-$ and $\bar{p}$ from
\cite{bib17} are also plotted, where the open triangles represent the 
$<p_T>$ obtained in pp interactions and the open circles correspond to
the $<p_T>$ measured in Au+Au collisions for different centralities. 
The $<p_T>$ of $\pi^-$, K$^-$ and $\bar{p}$ increases from peripheral 
to central collisions and from lighter to heavier particles. The 
$\pi^-$ $<p_T>$ increase from pp and most peripheral Au+Au collisions
to the most central Au+Au collisions is about 10$\%$. In the case of 
K$^-$ and $\bar{p}$, the increase is about 25$\%$ and 60$\%$, 
respectively. This behavior is expected from collective flow with 
hadronic rescattering. The $\phi$ meson has a very different behavior:
Its $<p_T>$ increases from pp collisions to Au+Au collisions. However,
there is little increase from peripherial Au+Au to central Au+Au
collisions although its mass is even higher than that of
$\bar{p}$. Some hadronic hydrodynamic calculations \cite{bib18}
predict collective flow expansion in proportion to particle mass, such
as those exhibited by $\pi^-$, $K^-$ and $\bar{p}$. The centrality
dependence of $<p_T>$ for the $\phi$ meson shows a distintive difference
from that for the $\pi^-$, $K^-$ and $\bar{p}$. This may indicate that
the evolution dynamics of the $\phi$ meson is less sensitive to hadronic 
rescattering.

\section{Conclusions}\label{concl}
We have presented rusults on $\phi$ meson production at mid-rapidity 
in Au+Au and pp collisions at $\sqrt{s_{NN}}$ = 200 GeV. The
spectrum in pp collisions is significantly different from that in
Au+Au collisions. The spectra in Au+Au collisions weakly depend
on the collision centralities. The extracted $\phi$ yield is a linear 
function of the number of charged particles produced in the
collision. The weak dependence of $\phi$ $<p_T>$ on the centralities in
Au+Au collisions is consistent with the expectation that $\phi$
does not interact strongly with nonstrange hadronic matter.
 
\section*{Note(s)}
\begin{notes}
\item[a]
E-mail: jgma@physics.ucla.edu
\end{notes}

\vfill\eject
\end{document}